\documentclass[12pt,,letterpaper]{JHEP3}
\usepackage{graphics}
\usepackage{epsfig}
\usepackage{slashed}

\title{Holographic non-relativistic fermionic fixed point by the charged dilatonic black hole}

\author{Wei-Jia Li\\
Department of Physics, Beijing Normal University, 100875 Beijing,
China\\
\email{wjli@mail.bnu.edu.cn}
}

\author{Ren\'e Meyer\\
Crete Center for Theoretical Physics, Department of Physics, \\
University of Crete, 71003 Heraklion, Greece\\
\email{meyer@physics.uoc.gr}}
\author{Hongbao Zhang\\
Crete Center for Theoretical Physics, Department of Physics, \\
University of Crete, 71003 Heraklion, Greece\\
\email{hzhang@physics.uoc.gr}
}

\abstract{Driven by the landscape of garden-variety condensed matter systems, we have investigated how the dual spectral function behaves at the non-relativistic as well as relativistic fermionic fixed point by considering the probe Dirac fermion in an extremal charged dilatonic black hole with zero entropy. Although the pattern for both of the appearance of flat band and emergence of Fermi surface is qualitatively similar to that given by the probe fermion in the extremal Reissner-Nordstrom AdS black hole,  we find a distinctly different low energy behavior around the  Fermi surface, which can be traced back to the different near horizon geometry. In particular,  with the peculiar near horizon geometry of our extremal charged dilatonic black hole, the low energy behavior exhibits the universal linear dispersion relation and scaling property, where the former indicates that the dual liquid is a Fermi one while the latter implies that the dual liquid is not exactly of Landau Fermi type.}
\preprint{CCTP-2011-38}
\begin{document}
\section{Introduction}
AdS/CFT correspondence, being a particular kind of strong/weak duality, has recently offered a rich playground for us to answer for those puzzling questions from condensed matter physics by means of the elegant black hole physics in one extra dimension. For a review of this vigorous subject, now dubbed as AdS/CMT, please refer to
 \cite{Hartnoll,Herzog1,McGreevy,Horowitz,Sachdev}.

Among others, with AdS/CFT correspondence, efforts have been made recently towards understanding the mysterious behaviors of existing non-Fermi liquids, which can not be well explained by the conventional approaches. Speaking specifically, with the fermionic correlator, which can be extracted at the boundary holographically by evolving the bulk Dirac equation in the  Reissner-Nordstrom AdS black hole with the ingoing boundary condition at the horizon, one can read off the Fermi surface with a rich spectrum of non-Fermi liquid behaviors\cite{Lee,LMV,CZS,FLMV1,FLMV2,FILMV,ILM,GM1}. However, as is well known, the Reissner-Nordstrom AdS black hole has a large ground state degeneracy, namely a finite entropy at zero temperature, which is somehow in tension with the third law of thermodynamics. Although it is believed that such a large ground state degeneracy is an artifact of the large $N$ approximation, it is favorable for one to evade this by instead playing with those black holes whose ground entropy vanishes. In addition, as a response to the man-made multiverse in condensed matter physics, it is also valuable to widen the range of boundary liquid behaviors by such a change of geometry in the bulk. 

With this in mind, some endeavors have been made to investigate the probe fermion in the desirable charged dilatonic black hole. In particular, taking into account the fact that the low energy behavior around Fermi surface is essentially controlled by the near horizon background, the authors in \cite{LKNT} have investigated various near horizon backgrounds, which can be characterized effectively by the two parameters $\beta$ and $\gamma$\footnote{With the two parameters $\beta$ and $\gamma$, the corresponding near horizon geometry goes like $ds^2\rightarrow -Cr^{2\gamma}dt^2+r^{2\beta}[(dx^1)^2+(dx^2)^2]+\frac{dr^2}{Cr^{2\gamma}}$ with $C$ a constant.}. Furthermore it is shown in \cite{LKNT} that the pattern for the low energy behavior around Fermi surface is determined by the value of $\beta+\gamma$. For $\beta+\gamma>1$ one gets the Fermi liquid behavior. For $\beta+\gamma<1$ there is no well-defined quasi-particle excitation. The case of $\beta+\gamma=1$ includes two branches. One is the near horizon background of the extremal Reissner-Nordstrom AdS black hole, where $\beta=0$ and $\gamma=1$. The other consists of those near horizon backgrounds for the extremal black holes with vanishing entropy. As a result,  this latter branch can give rise to the similar spectrum of non-Fermi liquid behaviors to that seen in the former branch in principle. However, some important quantities like Fermi momentum and Fermi velocity are not nailed down in \cite{LKNT}. In order to fix these quantities, it is better to have an exact full background solution with AdS as the UV completion, as it can make the involved numerical calculation easier. Actually, such an exact solution exists at least for the case of $\beta+\gamma=1$ with $\beta=\frac{1}{4}$ and $\gamma=\frac{3}{4}$\cite{GM2}. This has stimulated some works on the probe fermion in such a special background at various levels\cite{GR,Wu}.

On the other hand, as we know, in the context of AdS/CFT correspondence, not only such a modification in the bulk geometry, but also the change of boundary conditions can give rise to the different boundary field theory. In particular, as shown in \cite{LT1} very recently, in the latter manner, one can actually implement the holographic non-relativistic fermionic fixed points by imposing the Lorentz violating boundary condition rather than the Lorentz covariant one, which results in the presence of an infinite flat band in the boundary field theory. 

The purpose of this paper is to investigate how the fermionic correlator behaves at the non-relativistic fermionic fixed point by putting the probe fermion in the extremal charged dilatonic black hole with $\beta=\frac{1}{4}$ and $\gamma=\frac{3}{4}$ mentioned above. The rest of this paper is structured as follows. In the next section, we shall recall how both of the
relativistic and non-relativistic fermionic fixed points are
implemented by holography. Then we shall provide a concise review of the charged dilatonic black hole under consideration and its thermodynamics in the subsequent section.   In Section \ref{mainresult}, after
working out how to extract the fermionic
correlator by holography at the non-relativistic as well as relativistic fermionic fixed point in an explicit way, we shall present our
numerical results for the relevant quantities
associated with the fermionic correlator at the non-relativistic fermionic fixed point, where for comparison, the relevant results are also included for the fermionic correlator at the relativistic fermionic fixed point\footnote{The relevant calculations done in both \cite{GR} and \cite{Wu} are only for the probe fermion with the charge $q=2$ at the relativistic fermionic fixed point. As shown later, we have also sampled other values of $q$ in the current paper.}. Conclusions together with discussions will be addressed in
the end.

\section{Holographic implementation of various fermionic fixed points}
Let us start out by considering the bulk action for a probe Dirac fermion with the mass
$m$ and charge $q$
\begin{equation}
S_{bulk}=\int_\mathcal{M}
d^4x\sqrt{-g}i\bar{\psi}\Big[\frac{1}{2}(\overrightarrow{\slashed{D}}-
\overleftarrow{\slashed{D}})-m\Big]\psi
\end{equation}
in a general diagonal and rotationally invariant background, i.e.,
\begin{equation}
ds^2=-g_{tt}(r)dt^2+g_{rr}(r)dr^2+g_{xx}(r)[(dx^1)^2+(dx^2)^2], A_a=A_t(r)(dt)_a.
\end{equation}
Here $\bar{\psi}=\psi^\dag\Gamma^t$, and
$\overrightarrow{\slashed{D}}=(e_\mu)^a\Gamma^\mu[\partial_a+\frac{1}{4}[(\omega_{\rho\sigma})_a\Gamma^{\rho\sigma}-iqA_a]$, where $(e_\mu)^a$ form a set of orthogonal normal vector bases, and
Gamma matrices satisfy $\{\Gamma^\mu,\Gamma^\nu\}=2\eta^{\mu\nu}$
with the spin connection
$(\omega_{\mu\nu})_a=(e_\mu)_b\nabla_a(e_\nu)^b$ and 
$\Gamma^{\mu\nu}=\frac{1}{2}[\Gamma^\mu,\Gamma^\nu]$.  In what follows, we shall choose the orthogonal normal vector bases as 
\begin{equation}
(e_t)^a=\frac{1}{\sqrt{g_{tt}}}(\frac{\partial}{\partial
t})^a,(e_i)^a=\frac{1}{\sqrt{g_{xx}}}(\frac{\partial}{\partial
x^i})^a,(e_r)^a=\frac{1}{\sqrt{g_{rr}}}(\frac{\partial}{\partial
r})^a,
\end{equation}
whereby the non-vanishing components of spin connection can be
worked out as
\begin{equation}
(\omega_{tr})_a=-(\omega_{rt})_a=-\frac{\partial_r\sqrt{g_{tt}}}{\sqrt{g_{rr}}}(dt)_a,(\omega_{ir})_a=-(\omega_{ri})_a=\frac{\partial_r\sqrt{g_{xx}}}{\sqrt{g_{rr}}}(dx^i)_a.
\end{equation}
In addition, we would like to fix our convention of 
Gamma matrices as follows
\begin{eqnarray}
\Gamma^r=\left(
  \begin{array}{cc}
    -\sigma^3 & 0 \\
    0 & -\sigma^3 \\
  \end{array}
\right),\ \ \Gamma^t=\left(
             \begin{array}{cc}
               i\sigma^1 & 0 \\
               0 & i\sigma^1 \\
             \end{array}
           \right),\ \ \Gamma^1=\left(
             \begin{array}{cc}
               -\sigma^2 & 0 \\
               0 & \sigma^2 \\
             \end{array}
           \right),\ \ \Gamma^2=\left(
             \begin{array}{cc}
               0 & -i\sigma^2 \\
               i\sigma^2 & 0 \\
             \end{array}
           \right).\nonumber\\
\end{eqnarray}

As is well known, to have a well-defined variational principle for the above Dirac
action,  one must  add a boundary term, which turns out not to be unique, depending on the specific value of the mass parameter $m$\footnote{For simplicity but without loss
of generality, we shall focus on the case of $m\geq0$ in what
follows.}. To demonstrate this point, let us firstly perform the
variation of bulk action, which gives rise to
\begin{eqnarray}
\delta
S_{bulk}&=&i\int_\mathcal{M}d^4x\sqrt{-g}\Big[\delta\bar{\psi}(\overrightarrow{\slashed{D}}-m)\psi-\bar{\psi}(\overleftarrow{\slashed{D}}+m)\delta\psi\Big]\nonumber\\
&&+\frac{i}{2}\int_{\partial\mathcal{M}}d^3x
\sqrt{-h}(\bar{\psi}_-\delta\psi_+-\bar{\psi}_+\delta\psi_-+\delta\bar{\psi}_+\psi_--\delta\bar{\psi}_-\psi_+)
\end{eqnarray}
where $h=\frac{g}{g_{rr}}$ is the determinant of induced metric on
the boundary, and $\psi_\pm=\frac{1}{2}(1\pm\Gamma^r)\psi$. Note
that the bulk Dirac equation
\begin{equation}\label{Dirac}
(\overrightarrow{\slashed{D}}-m)\psi=0
\end{equation}
is first order. So one is not allowed to nail down all components of
$\psi$ simultaneously. Instead one could 
 fix merely half of the components of $\psi$, which can
actually be carried out by adding various boundary terms.  Now let us go through these viable boundary terms one by one.

The conventional boundary term to be chosen is the Lorentz
covariant one, i.e.,
\begin{equation}
S_{bdy}=\frac{i}{2}\int_{\partial\mathcal{M}}\sqrt{-h}\bar{\psi}\psi=\frac{i}{2}\int_{\partial\mathcal{M}}\sqrt{-h}(\bar{\psi}_-\psi_++\bar{\psi}_+\psi_-),
\end{equation}
whereby the variation of the full on-shell action reads
\begin{equation}
\delta S_{bulk}+\delta
S_{bdy}=i\int_{\partial\mathcal{M}}d^3x\sqrt{-h}(\bar{\psi}_-\delta\psi_++\delta\bar{\psi}_+\psi_-),
\end{equation}
which apparently vanishes if the Dirichlet boundary
condition is imposed on $\psi_+$. This choice of boundary
condition is usually referred to as the standard quantization for the
Dirac field. If one expresses $\psi$ together with $\psi_+$ and $\psi_-$ as

\begin{eqnarray}
\psi=(-h)^{-\frac{1}{4}}\left(
  \begin{array}{c}
    y_1 \\
    z_1 \\
    y_2 \\
    z_2 \\
  \end{array}
\right),\ \, \psi_+=(-h)^{-\frac{1}{4}}\left(
  \begin{array}{c}
    0 \\
    z_1 \\
    0 \\
    z_2 \\
  \end{array}
\right),\ \, \ \psi_-=(-h)^{-\frac{1}{4}}\left(
  \begin{array}{c}
    y_1 \\
    0 \\
    y_2 \\
    0 \\
  \end{array}
\right),
\end{eqnarray}
then the above formula can be reduced to
\begin{equation}\label{standard}
\delta S_{bulk}+\delta
S_{bdy}=-\int_{\partial M}d^3x(y^\dagger_1\delta z_1+y^\dagger_2\delta z_2+\delta z^\dagger_1y_1+\delta z^\dagger_2y_2).
\end{equation}
Accordingly, the dual boundary field theory is a
Lorentz covariant CFT where loosely speaking $z$s are identified as the
fermionic sources, and the corresponding dual operators are given by $y$s with
dimension
\begin{equation}
\Delta[y_1]=\Delta[y_2]=\frac{3}{2}+m.
\end{equation}
When the mass parameter falls into the window $0\leq
m<\frac{1}{2}$, then by normalizablity one can have other boundary conditions to choose. The
first one is the so called alternative quantization, which can be
realized simply by adding the boundary term with opposite sign,
i.e.,
\begin{equation}
S_{bdy}=-\frac{i}{2}\int_{\partial\mathcal{M}}\sqrt{-h}\bar{\psi}\psi=-\frac{i}{2}\int_{\partial\mathcal{M}}\sqrt{-h}(\bar{\psi}_-\psi_++\bar{\psi}_+\psi_-).
\end{equation}
Whence the variation of full on-shell action can be obtained as
\begin{eqnarray}\label{alternative}
\delta S_{bulk}+\delta
S_{bdy}&=&-i\int_{\partial\mathcal{M}}d^3x
\sqrt{-h}(\bar{\psi}_+\delta\psi_-+\delta\bar{\psi}_-\psi_+)\nonumber\\
&=&\int_{\partial M}d^3x(z^\dagger_1\delta y_1+z^\dagger_2\delta y_2+\delta y^\dagger_1z_1+\delta y^\dagger_2z_2).
\end{eqnarray}
This now gives rise to a well defined variational principle 
if one imposes the Dirichlet boundary condition on $\psi_-$. The
dual boundary field theory is still a Lorentz covariant CFT with the
fermionic sources and dual operators swapped. The dimension of
operators is thus given by
\begin{equation}
\Delta[z_1]=\Delta[z_2]=\frac{3}{2}-m.
\end{equation}
We still have other boundary conditions to play with if we do not restrict ourselves onto
 the Lorentz covariant boundary field theory. In
particular, as demonstrated in \cite{LT1}, a non-relativistic fermionic fixed
point can be implemented by adding the following
boundary term, i.e.,
\begin{equation}\label{lvb1}
S_{bdy}=\frac{1}{2}\int_{\partial\mathcal
{M}}d^3x\sqrt{-h}\bar{\psi}\Gamma^1\Gamma^2\psi,
\end{equation}
which obviously breaks the Lorentz covariance, but still preserves the U(1) global symmetry under $\psi\rightarrow e^{i\theta}\psi$, rotation and scale invariance.
To see what both of the fermionic sources
and dual operators look like in this case,   
let us combine the explicit expression for the variation of bulk on-shell action
\begin{equation}
\delta S_{bulk}=-\frac{1}{2}\int_{\partial\mathcal {M}}d^3x(\delta
z^\dag_1y_1+\delta z^\dag_2y_2-\delta y^\dag_1z_1-\delta
y^\dag_2z_2-z^\dag_1\delta y_1-z^\dag_2\delta y_2+ y^\dag_1\delta
z_1+y^\dag_2\delta z_2)
\end{equation}
with the variation of  the Lorentz violating
boundary term 
\begin{equation}
\delta S_{bdy}=-\frac{1}{2}\int_{\partial\mathcal
{M}}d^3x\delta(z^\dag_2y_1+y^\dag_2z_1+z^\dag_1y_2+y^\dag_1z_2).
\end{equation}
As a result, the variation of full on-shell action can be massaged as
\begin{eqnarray}\label{non1}
\delta S_{bulk}+\delta S_{bdy}&=&-\frac{1}{2}\int_{\partial\mathcal
{M}}d^3x
[\delta(z^\dag_1+z^\dag_2)(y_1+y_2)+\delta(y^\dag_1-y^\dag_2)(z_2-z_1)\nonumber\\
&&+(z^\dag_2-z^\dag_1)\delta(y_1-y_2)+(y^\dag_1+y^\dag_2)\delta(z_1+z_2)]\nonumber\\
&=& -\int_{\partial\mathcal {M}}d^3x (\delta Z^\dag_1
Y_1+Z^\dag_2\delta Y_2+Y^\dag_1\delta Z_1+\delta Y^\dag_2 Z_2),
\end{eqnarray}
where we have defined
$(Y_1,Y_2)=\frac{1}{\sqrt{2}}(y_1+y_2,y_1-y_2)$, and
$(Z_1,Z_2)=\frac{1}{\sqrt{2}}(z_1+z_2,z_2-z_1)$. Hence we can read off the
fermionic sources and dual operators as $(Z_1,Y_2)$
and $(Y_1,Z_2)$ respectively with  the dimension of operators given by
\begin{equation}
\Delta[Y_1]=\frac{3}{2}+m,\Delta[Z_2]=\frac{3}{2}-m.
\end{equation}
Similarly, another non-relativistic fermionic fixed point can be easily realized by adding the boundary term opposite to the above Lorentz violating one, i.e.,
\begin{equation}\label{lvb2}
S_{bdy}=-\frac{1}{2}\int_{\partial\mathcal
{M}}d^3x\sqrt{-h}\bar{\psi}\Gamma^1\Gamma^2\psi,
\end{equation}
which eventually leads to the variation of full on-shell action as
\begin{equation}\label{non2}
\delta S_{bulk}+\delta S_{bdy}= \int_{\partial\mathcal {M}}d^3x (Z^\dag_1
\delta Y_1+\delta Z^\dag_2 Y_2+\delta Y^\dag_1 Z_1+Y^\dag_2\delta Z_2).
\end{equation}
This implies that the corresponding fermionic sources and dual operators are interchanged with the dimension of operators given by
\begin{equation}
\Delta[Z_1]=\frac{3}{2}-m, \Delta[Y_2]=\frac{3}{2}+m.
\end{equation}
Furthermore, more boundary terms are viable if one does not stick to the U(1) global symmetry.
For more details, please refer to \cite{LT1,LT2}. 

\section{The charged dilatonic black hole and its thermodynamics}
Start with the Einstein-Maxwell-dilaton background 
\begin{equation}\label{blackhole}
ds^2=e^{2B}[-fdt^2+(dx^1)^2+(dx^2)^2]+\frac{1}{e^{2B}}\frac{dr^2}{f}, A_a=\Phi(dt)_a, \phi=\frac{1}{2}\ln(1+\frac{Q}{r})
\end{equation}
where
\begin{equation}
B=\ln\frac{r}{L}+\frac{3}{4}\ln(1+\frac{Q}{r}), f=1-\frac{\nu L^2}{(Q+r)^3}, \Phi=\frac{\sqrt{3Q\nu}}{Q+r}-\frac{\sqrt{3Q}\nu^\frac{1}{6}}{L^\frac{2}{3}}.
\end{equation}
This charged dilatonic black hole can be regarded as a solution to the following action, i.e.,
\begin{equation}
S=\frac{1}{2\kappa^2}\int_Md^4x\sqrt{-g}[R-\frac{1}{4}e^\phi F_{ab}F^{ab}-\frac{3}{2}\nabla_a\phi\nabla^a\phi+\frac{6}{L^2}\cosh\phi],
\end{equation}
where $\kappa^2=8\pi G$ with $G$ the Newton gravitational constant, $R$ is the Ricci scalar, and $F=dA$ is the field strength\cite{GM2}\footnote{For its embedding into string theory and corresponding thermodynamic instability, please refer to \cite{GR}.} . By regularizing the conical singularity in the corresponding Euclidean sector, the temperature of black hole can be obtained as
\begin{equation}
T=\frac{1}{4\pi}\frac{\partial_rg_{tt}}{\sqrt{g_{tt}g_{rr}}}|_{r_H}=\frac{3\nu^\frac{1}{6}}{4\pi L^\frac{5}{3}}\sqrt{r_H},
\end{equation}
where the horizon is located at the place where $f$ vanishes, namely $r_H=\nu^\frac{1}{3}L^\frac{2}{3}-Q$. Thus the zero temperature can be achieved by setting $\nu=\frac{Q^3}{L^2}$. In addition, by the Bekenstein-Hawking formula, the entropy density reads
\begin{equation}
s=\frac{2\pi\nu^\frac{1}{2}}{\kappa^2L}\sqrt{r_H}=\frac{8\pi^2L^\frac{2}{3}\nu^\frac{1}{3}}{\kappa^2}T,
\end{equation}
which obviously vanishes at zero temperature as we desire.
On the other hand,  note that the above spacetime is asymptotically AdS, i.e.,
\begin{equation}
ds^2\rightarrow \frac{r^2}{L^2}[-dt^2+(dx^1)^2+(dx^2)^2]+\frac{L^2}{r^2}dr^2; r\rightarrow\infty.
\end{equation}
So by the holographic dictionary,  the bulk gauge field $A_t$ evaluated at the boundary serves as the chemical potential, i.e., 
\begin{equation}
\mu=-\frac{\sqrt{3Q}\nu^\frac{1}{6}}{L^\frac{2}{3}},
\end{equation}
and the corresponding dual charge density can be obtained by the variation of the on-shell action as
\begin{equation}
\rho=\lim_{r\rightarrow\infty}\frac{\delta S}{\delta A_t}=-\frac{\sqrt{3Q\nu}}{2\kappa^2L^2}.
\end{equation}
Whence in the low temperature limit, the entropy density can be approximated as
\begin{equation}
s\approx-\frac{8\sqrt{3}\pi^3L}{3\kappa^2}\mu T\approx\frac{8\sqrt{2}\pi^2L^\frac{3}{2}}{3^\frac{1}{4}\kappa}(-\rho)^\frac{1}{2}T.
\end{equation}
Therefore when one goes to zero temperature, the specific heats at constant charge density and constant chemical potential, namely
$c_\rho=T(\frac{\partial s}{\partial T})_\rho$ and $c_\mu=T(\frac{\partial s}{\partial T})_\mu$ are both linear with respect to the temperature, which is reminiscent of the Landau Fermi liquid behavior.

In what follows we shall work exclusively with the extremal charged dilatonic black hole, where for convenience we set $\nu=Q=L=1$.

\section{Holographic non-relativistic fermion by charged dilatonic black hole}\label{mainresult}
\subsection{Holographic setup}
Generically, most of the relevant information regarding the fermionic system
can be read out of its single particle fermionic correlator, namely
the retarded Green function $G_R$. Now we shall show how such a retarded Green
function can be worked out by holography.

To proceed, we would like to start by setting $\psi=(-h)^{-\frac{1}{4}}\varphi$, then the bulk Dirac
equation can be written as
\begin{equation}
\frac{\Gamma^r\partial_r\varphi}{\sqrt{g_{rr}}}+\frac{\Gamma^t(\partial_t-iqA_t)\varphi}{\sqrt{g_{tt}}}+\frac{\Gamma^i\partial_i\varphi}{\sqrt{g_{xx}}}
-m\varphi=0.
\end{equation}
By the rotation symmetry in the spatial directions, without loss of
generality, we shall let $\varphi=e^{-i\omega
t+ikx^1}\tilde{\varphi}$,  which gives rise to the Dirac equation as
\begin{equation}
\frac{\sqrt{g_{xx}}}{\sqrt{g_{rr}}}(\Gamma^r\partial_r-m\sqrt{g_{rr}})\tilde{\varphi}+(-iu\Gamma^t+ik\Gamma^1)\tilde{\varphi}=0,
\end{equation}
where
\begin{equation}
u=\frac{\sqrt{g_{xx}}}{\sqrt{g_{tt}}}(\omega+qA_t).
\end{equation}
Next set $\tilde{\varphi}=\left(
                       \begin{array}{c}
                         \tilde{\varphi}_1 \\
                         \tilde{\varphi}_2\\
                       \end{array}
                     \right)$,
                     then with our representation of Gamma
                     matrices, the equation of motion can be further simplified as
\begin{equation}
\frac{\sqrt{g_{xx}}}{\sqrt{g_{rr}}}(\partial_r+m\sqrt{g_{rr}}\sigma_3)\tilde{\varphi}_I=[i\sigma_2u+(-1)^I
k\sigma_1]\tilde{\varphi}_I,
\end{equation}
with $I=1,2$. Furthermore, by $\tilde{\varphi}_I=\left(
                     \begin{array}{c}
                       \tilde{y}_I \\
                       \tilde{z}_I \\
                     \end{array}
                   \right)$, the above equation of motion leads to the following flow equation, i.e.,

                   \begin{equation}
                   \frac{\sqrt{g_{xx}}}{\sqrt{g_{rr}}}\partial_r\xi_I=-2m\sqrt{g_{xx}}\xi_I+[u+(-1)^Ik]+[u-(-1)^Ik]\xi_I^2
                   \end{equation}
where $\xi_I=\frac{\tilde{y}_I}{\tilde{z}_I}$. Now substitute the
particular background (\ref{blackhole}) into the above equations, we
wind up with
\begin{equation}\label{eom}
(e^{2B}\sqrt{f}\partial_r+me^{B}\sigma_3)\tilde{\varphi}_I=[\frac{i\sigma_2}{\sqrt{f}}(\omega+q\Phi)+(-1)^Ik\sigma_1]\tilde{\varphi}_I
\end{equation}
for the equation of motion, and
\begin{equation}\label{flow}
e^{2B}\sqrt{f}\partial_r\xi_I=-2me^{B}\xi_I+[\frac{1}{\sqrt{f}}(\omega+q\Phi)+(-1)^Ik]+[\frac{1}{\sqrt{f}}(\omega+q\Phi)-(-1)^Ik]\xi_I^2
\end{equation}
for the flow equation. Whence near the boundary, namely when $r$ approaches the infinity,
$\tilde{\varphi}_I$ take the following asymptotic behavior, i.e.,
\begin{equation}
\tilde{\varphi}_I\rightarrow c_I r^m\left(
                                                  \begin{array}{c}
                                                    0 \\
                                                    1 \\
                                                  \end{array}
                                                \right)+d_I
                                                r^{-m}\left(
                                                        \begin{array}{c}
                                                          1 \\
                                                          0 \\
                                                        \end{array}
                                                      \right).
\end{equation}
The ratio $G_I=\frac{d_I}{c_I}$ can be determined by imposing the
in-going boundary condition for $\tilde{\varphi}_I$ at the horizon, where $\tilde{\varphi}_I$ behave as
\begin{equation}
\tilde{\varphi}_I\propto \left(
                                                  \begin{array}{c}
                                                    i \\
                                                    1 \\
                                                  \end{array}
                                                \right)e^{-i\omega
                                                r_*}
 \end{equation}
 for $\omega\neq 0$, and
 
 \begin{equation}
 \tilde{\varphi}_I\propto \left(
                                                  \begin{array}{c}
                                                    |k|\\
                                                    (-1)^Ik\\
                                                  \end{array}
                                                \right)e^{|k|
                                                r^*} \end{equation}
 for $\omega=0$, where
 $r_*=\int\frac{dr}{e^{2B}f}$ and $r^*=\int\frac{dr}{e^{2B}\sqrt{f}}$\footnote{This  can be obtained by looking at the equation of motion (\ref{eom}) near the horizon, where the equation of motion can be approximated as
$e^{2B}\sqrt{f}\partial_r\tilde{\varphi}_I=\frac{i\sigma_2}{\sqrt{f}}\omega\tilde{\varphi}_I$ for $\omega\neq0$ and $e^{2B}\sqrt{f}\partial_r\tilde{\varphi}_I=(-1)^Ik\sigma_1\tilde{\varphi}_I
$ for $\omega=0$ respectively.}. Alternatively, this ratio can also be worked out in a more convenient way as
                               \begin{equation}
                               G_I=\lim_{r\rightarrow\infty}r^{2m}\xi_I
                               \end{equation}
 by solving the flow equation (\ref{flow}) with the boundary condition at the horizon
 \begin{equation}\label{nonzero}
 \xi_I=i
 \end{equation}
 when $\omega\neq 0$ and
 \begin{equation}\label{zero}
 \xi_I=(-1)^I\mathtt{sign}(k)
 \end{equation}
 when $\omega=0$. It is noteworthy that distinct from the case for the probe fermion in the extremal Reissner-Nordstrom AdS black hole, where the boundary condition is not always real, and dependent on the specific value of $m$, $q$, and $k$\cite{LMV}, the above boundary condition for $\omega=0$ depends solely on the sign of $k$. Such an intriguing boundary condition comes essentially from the peculiar near horizon background of our extremal charged dilatonic black hole, i.e.,
 \begin{equation}
 ds^2\rightarrow-3r^\frac{3}{2}dt^2+r^\frac{1}{2}[(dx^1)^2+(dx^2)^2]+\frac{dr^2}{3r^\frac{3}{2}} , \Phi\rightarrow-\sqrt{3}r; r\rightarrow 0,
 \end{equation}
 where the metric  is neither the AdS type nor the Lifshitz like\footnote{As expected, one can read off $\beta=\frac{1}{4}$ and $\gamma=\frac{3}{4}$ from the asymptotic form of metric.}.  
 
 Now by the recipe of AdS/CFT, the retarded fermionic Green function is defined through the variation of
full on-shell action (\ref{standard}) and (\ref{alternative}) as
\begin{eqnarray}
\left(
               \begin{array}{c}
                 d_1 \\
                 d_2 \\
               \end{array}
             \right)
=G_R\left(
               \begin{array}{c}
                 c_1 \\
                 c_2 \\
               \end{array}
             \right)
\end{eqnarray}
for the standard quantization and
\begin{eqnarray}
\left(
               \begin{array}{c}
                 c_1 \\
                 c_2 \\
               \end{array}
             \right)
=G_R\left(
               \begin{array}{c}
                 -d_1 \\
                 -d_2 \\
               \end{array}
             \right)
             \end{eqnarray}
for the alternative quantization respectively. Whence one can easily read off the retarded fermionic Green function as
\begin{eqnarray}\label{sg}
G_R=\left(
                                 \begin{array}{cc}
                                     \frac{d_1}{c_1} & 0 \\
                                     0 & \frac{d_2}{c_2} \\
                                   \end{array}
                                 \right)=\left(
                                 \begin{array}{cc}
                                     G_1 & 0 \\
                                     0 & G_2 \\
                                   \end{array}
                                 \right)
                                 \end{eqnarray}
for the standard quantization and
\begin{eqnarray}\label{ag}
G_R=\left(
                                 \begin{array}{cc}
                                     -\frac{c_1}{d_1} & 0 \\
                                     0 & -\frac{c_2}{d_2} \\
                                   \end{array}
                                 \right)=\left(
                                 \begin{array}{cc}
                                     -\frac{1}{G_1} & 0 \\
                                     0 & -\frac{1}{G_2} \\
                                   \end{array}
                                 \right)
                                 \end{eqnarray}
for the alternative quantization respectively.  It is a little bit cumbersome to work out the retarded Green function for the two non-relativistic fermionic fixed points, which is our present task. It follows from (\ref{non1}) that the corresponding retarded Green function is defined through the following relation, i.e., 
\begin{eqnarray}
\left(
               \begin{array}{c}
                 D_1 \\
                 C_2 \\
               \end{array}
             \right)
=G_R\left(
               \begin{array}{c}
                 C_1 \\
                 D_2 \\
               \end{array}
             \right)=\left(
                                 \begin{array}{cc}
                                     \alpha & \tau \\
                                     \xi & \eta \\
                                   \end{array}
                                 \right)
\left(
               \begin{array}{c}
                 C_1 \\
                 D_2 \\
               \end{array}
             \right),
\end{eqnarray}
where $(D_1,D_2)=\frac{1}{\sqrt{2}}(d_1+d_2,d_1-d_2)$ and
$(C_1,C_2)=\frac{1}{\sqrt{2}}(c_1+c_2,c_2-c_1)$. To be more explicit,
from such a relation, we have
\begin{eqnarray}
d_1+d_2=(\alpha\frac{c_1}{d_1}+\tau)d_1+(\alpha\frac{c_2}{d_2}-\tau)d_2,\nonumber\\
\frac{c_2}{d_2}d_2-\frac{c_1}{d_1}d_1=(\xi\frac{c_1}{d_1}+\eta)d_1+(\xi\frac{c_2}{d_2}-\eta)d_2.
\end{eqnarray}
Taking into account that $d_1$ and $d_2$ are independent of each other, we end up with
\begin{eqnarray}
\frac{\alpha}{G_1}+\tau=1,\ \ \ \frac{\alpha}{G_2}-\tau=1,\nonumber\\
\xi+\eta G_1=-1,\ \ \ \ \xi-\eta G_2=1,
\end{eqnarray}
which yields $G_R$ in terms of $G_I$
\begin{eqnarray}\label{ng1}
G_R=\left(
               \begin{array}{cc}
                 \frac{2G_1G_2}{G_1+G_2} & \frac{G_1-G_2}{G_1+G_2} \\
                 \frac{G_1-G_2}{G_1+G_2} & \frac{-2}{G_1+G_2} \\
               \end{array}
             \right).
\end{eqnarray}
 As a result, $\mathtt{det}(G_R)=-1$,  and the eigenvalues of $G_R$  as well as the trace can be
 worked out as
\begin{eqnarray}\label{e2}
&&\lambda_\pm=\frac{G_1G_2-1\pm\sqrt{1+G_1^2+G_2^2+G_1^2G_2^2}}{G_1+G_2}\\
&&\mathtt{Tr}(G_R)=\lambda_++\lambda_-=\frac{2G_1G_2-2}{G_1+G_2}.
\end{eqnarray}
Similarly, according to (\ref{non2}),  one can manipulate the following relation, i.e.,

\begin{eqnarray}
\left(
               \begin{array}{c}
                 C_1 \\
                 D_2 \\
               \end{array}
             \right)
=G_R\left(
               \begin{array}{c}
                 -D_1 \\
                 -C_2 \\
               \end{array}
             \right),
\end{eqnarray}
to extract the retarded Green function as
\begin{eqnarray}\label{ng1}
G_R=\left(
               \begin{array}{cc}
                 \frac{-2}{G_1+G_2} & \frac{G_2-G_1}{G_1+G_2} \\
                 \frac{G_2-G_1}{G_1+G_2} & \frac{2G_1G_2}{G_1+G_2} \\
               \end{array}
             \right)
\end{eqnarray}
 for another non-relativistic fermionic fixed point, which has the same eigenvalues and trace as those for the former non-relativistic fermionic fixed point. With hindsight,  this is reasonable as these two non-relativistic fermionic fixed points are connected by the boundary parity transformation\cite{LT1}. So in what follows we shall not distinguish these two non-relativistic fermionic fixed points any more. 

 It follows from
the flow equation (\ref{flow}) together with the boundary condition (\ref{nonzero}) and (\ref{zero}) that $G_1$ and $G_2$ are related to
each other as
\begin{equation}
G_2(\omega,k)=G_1(\omega,-k).
\end{equation}
Therefore both of the trace and eigenvalues of our retarded Green
functions are invariant under the transformation $k\rightarrow -k$ as
should be the case due to the rotation symmetry mentioned above.
Furthermore, we can also have
\begin{equation}
G_1(\omega,k;q)=-G_2^*(-\omega,k;-q).
\end{equation}
So it is essentially enough to restrict ourselves to non-negative
$k$ and $q$ .  To make our life easier,  below we shall
work exclusively with the case of $m=0$, where with the flow equation as well as the boundary condition, we can further have
\begin{equation}\label{easy}
G_1(\omega,k)=-\frac{1}{G_2(\omega,k)},
\end{equation}
which means that the alternative quantization is equivalent to the standard one for the case of $m=0$. 
In addition, the two eigenvalues can thus be further simplified as
\begin{equation}
 \lambda_+=\frac{G_1-1}{G_1+1} , \lambda_-=\frac{1+G_1}{1-G_1}
  \end{equation}
for the non-relativistic fermionic fixed point.
\subsection{Numerical results}
As mentioned above, we shall focus exclusively on the massless probe fermion in the zero temperature soup, where the chemical potential felt by the probe fermion is given by $\Omega=-\sqrt{3}q$. For comparison, below we shall also include the relevant numerical results for the relativistic fermionic fixed point.
\begin{figure}[ht]
\centering      
\includegraphics[width=.48\textwidth]{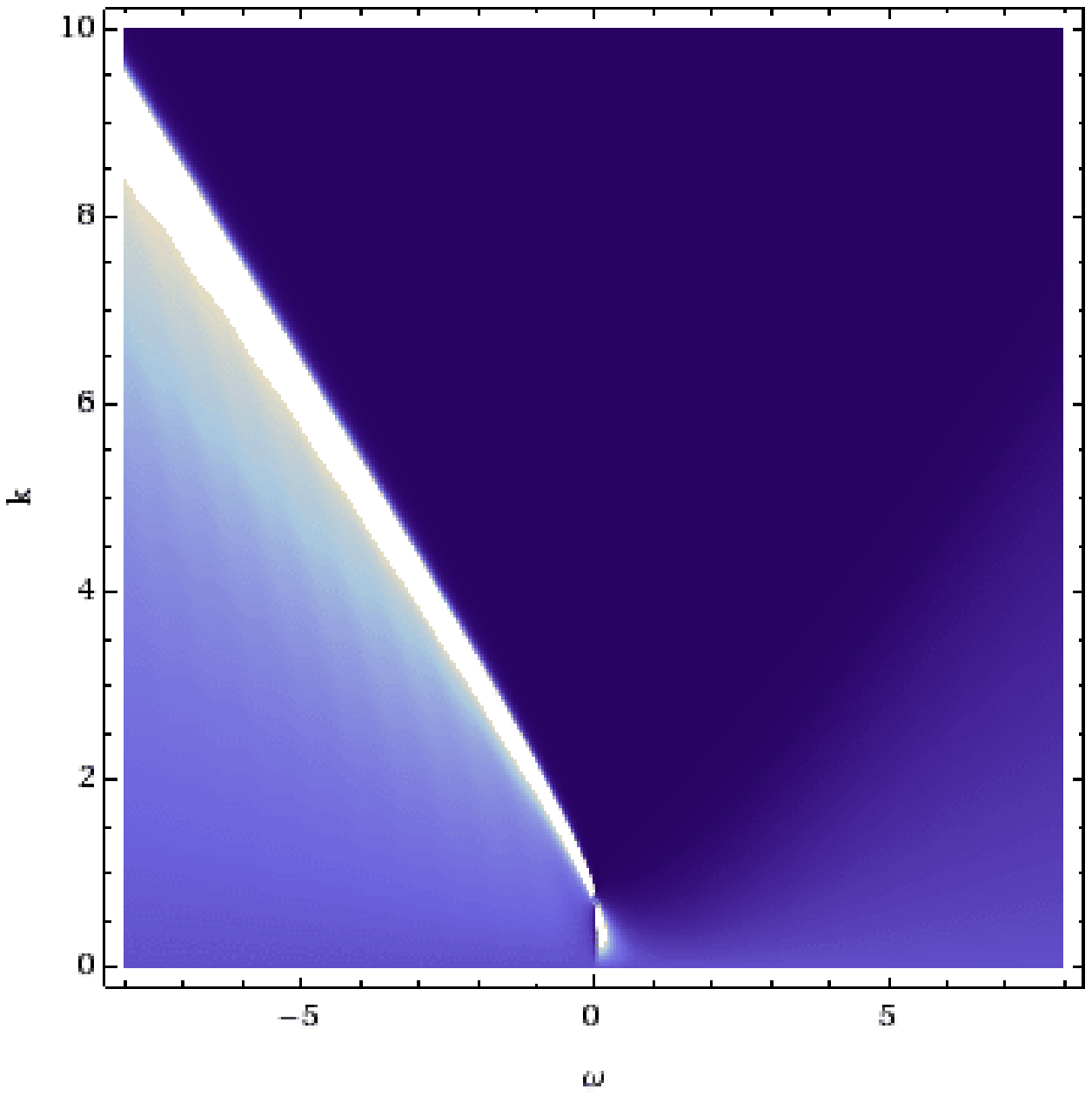}
      \includegraphics[width=.48\textwidth]{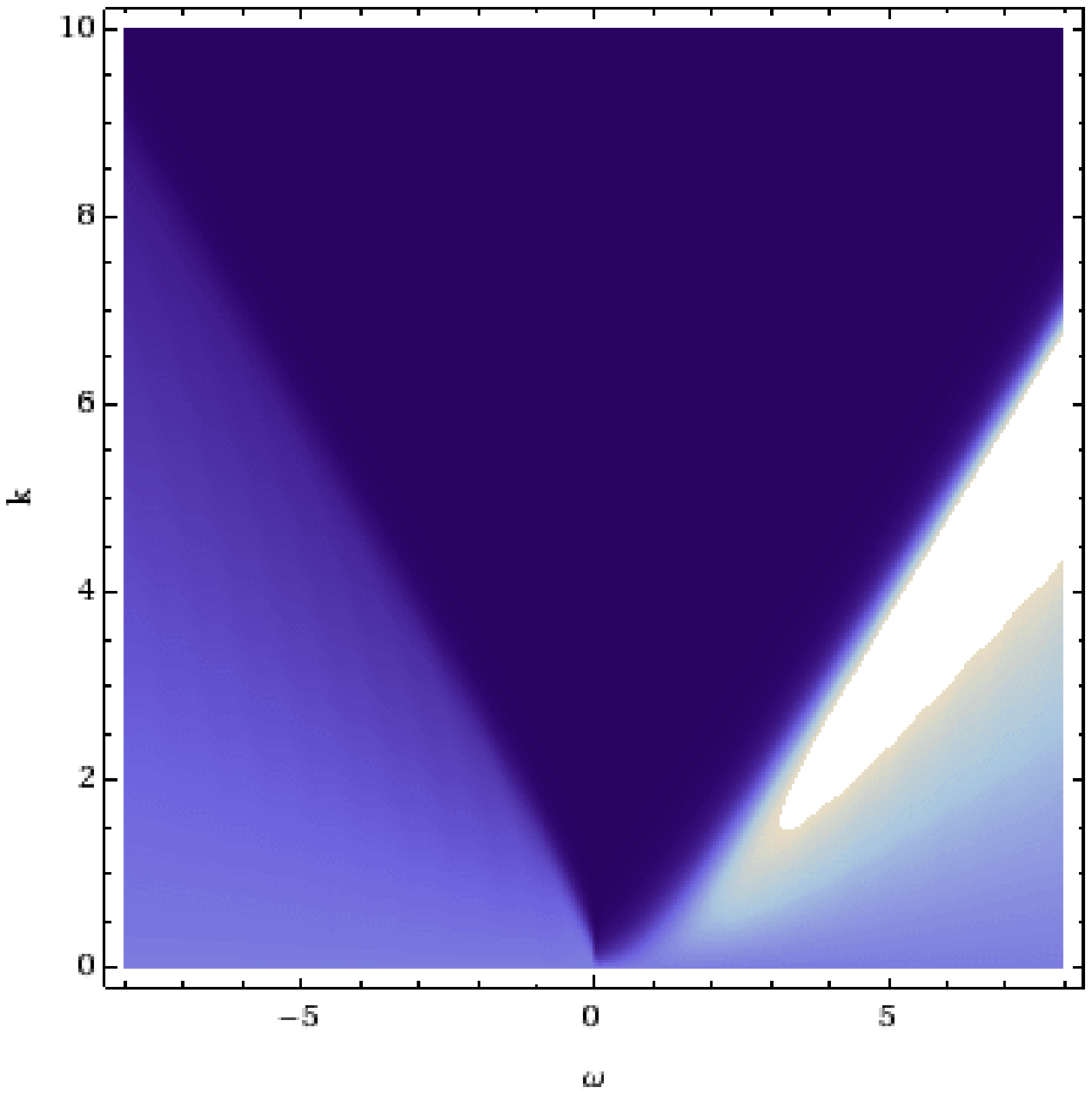}   
        \includegraphics[width=.48\textwidth]{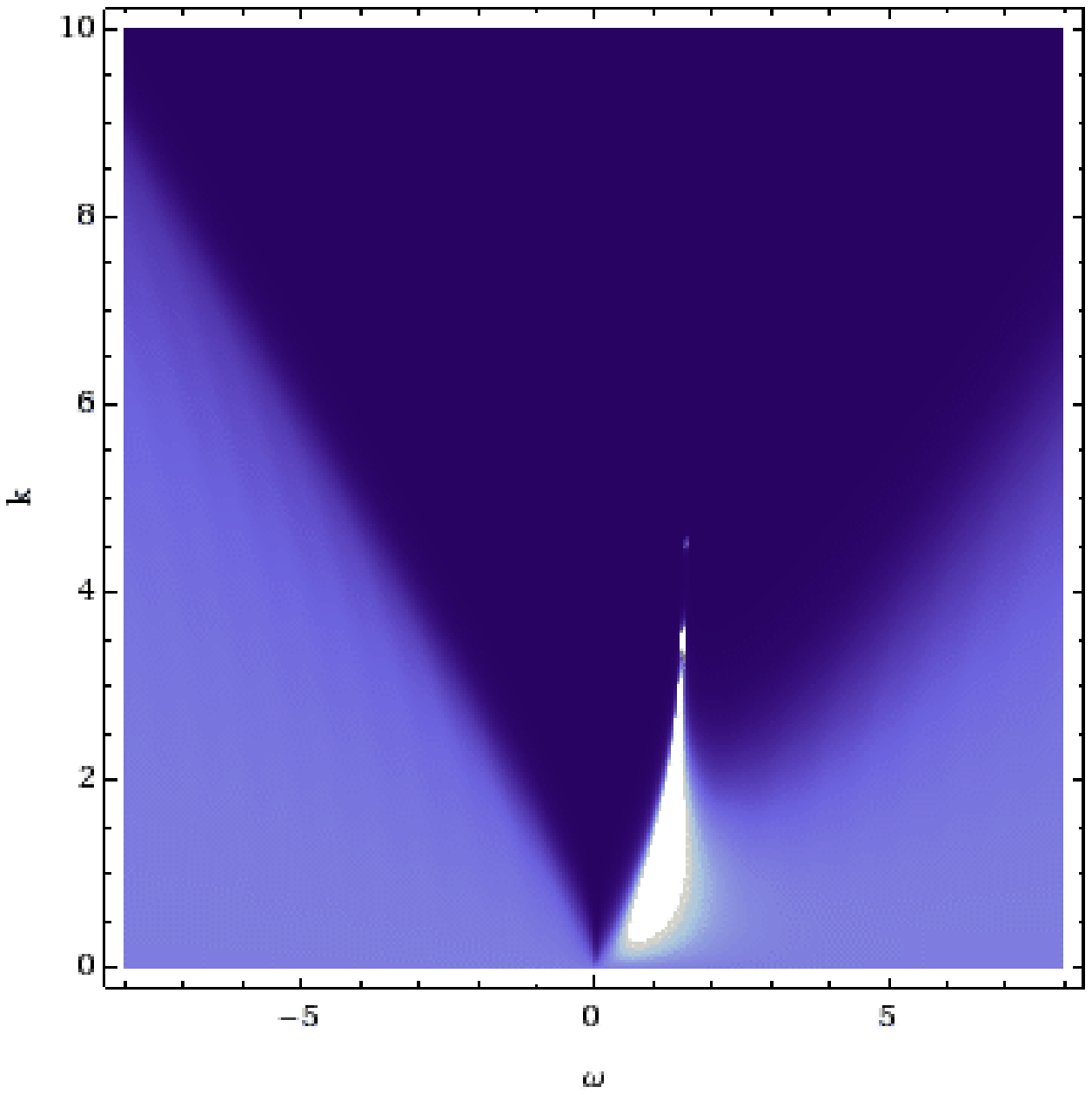}
      \includegraphics[width=.48\textwidth]{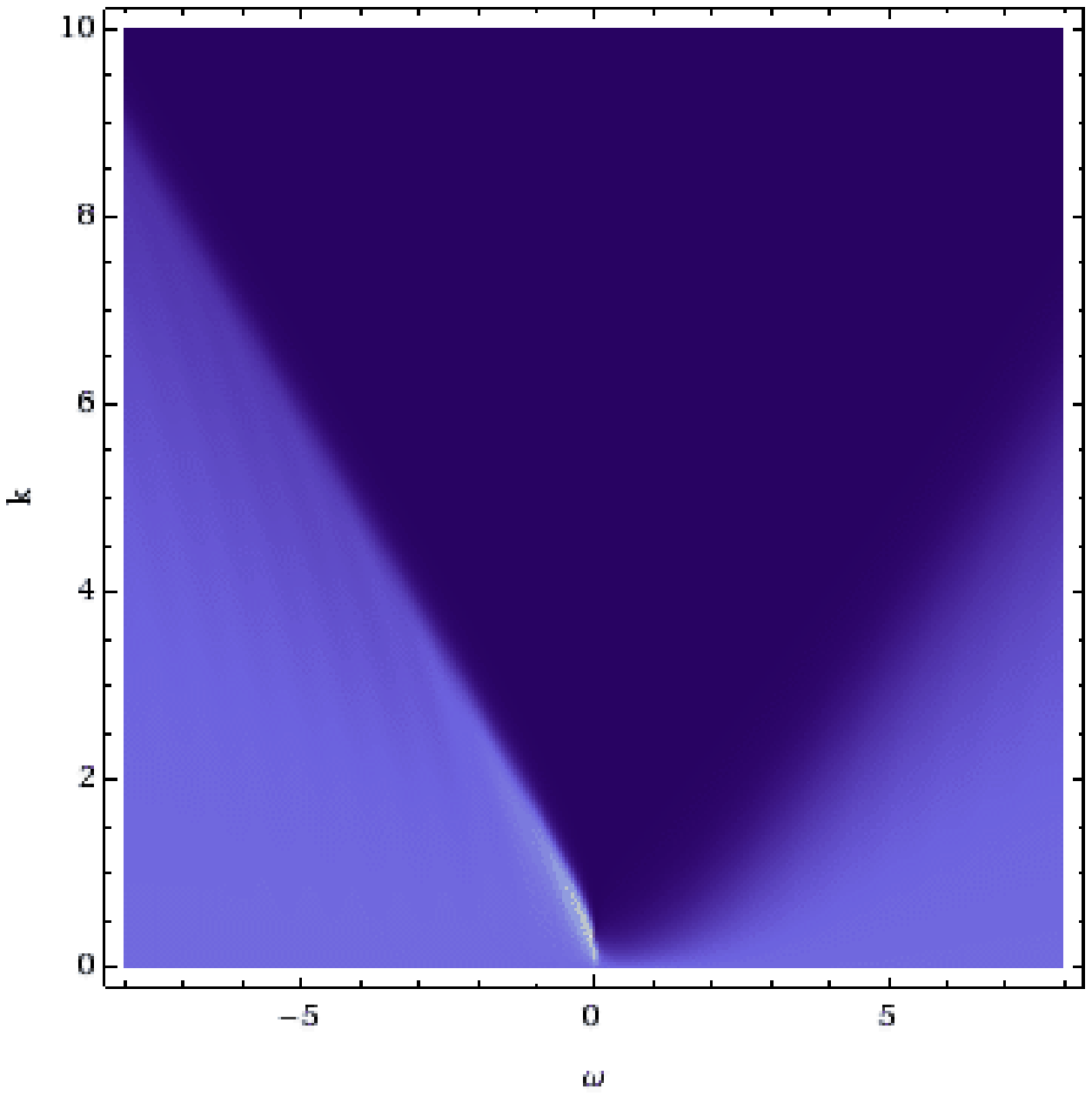}        \caption{The density plots of spectral function for the case of $q=1$, where the upstairs and downstairs plots are for the relativistic  and non-relativistic fermionic fixed point respectively. }
       \label{q1}
\end{figure}

\begin{figure}[ht]
\centering      
\includegraphics[width=.48\textwidth]{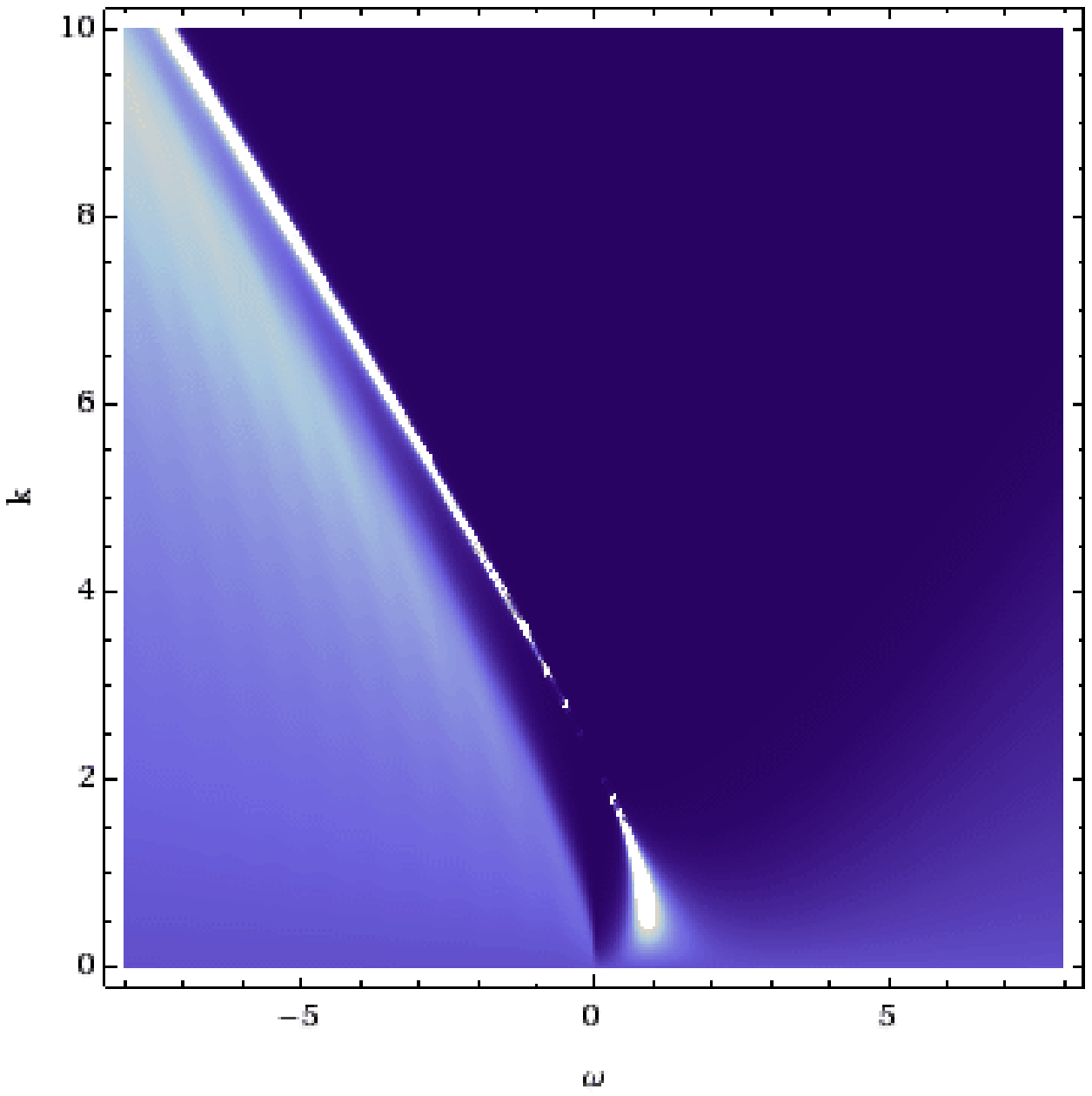}
      \includegraphics[width=.48\textwidth]{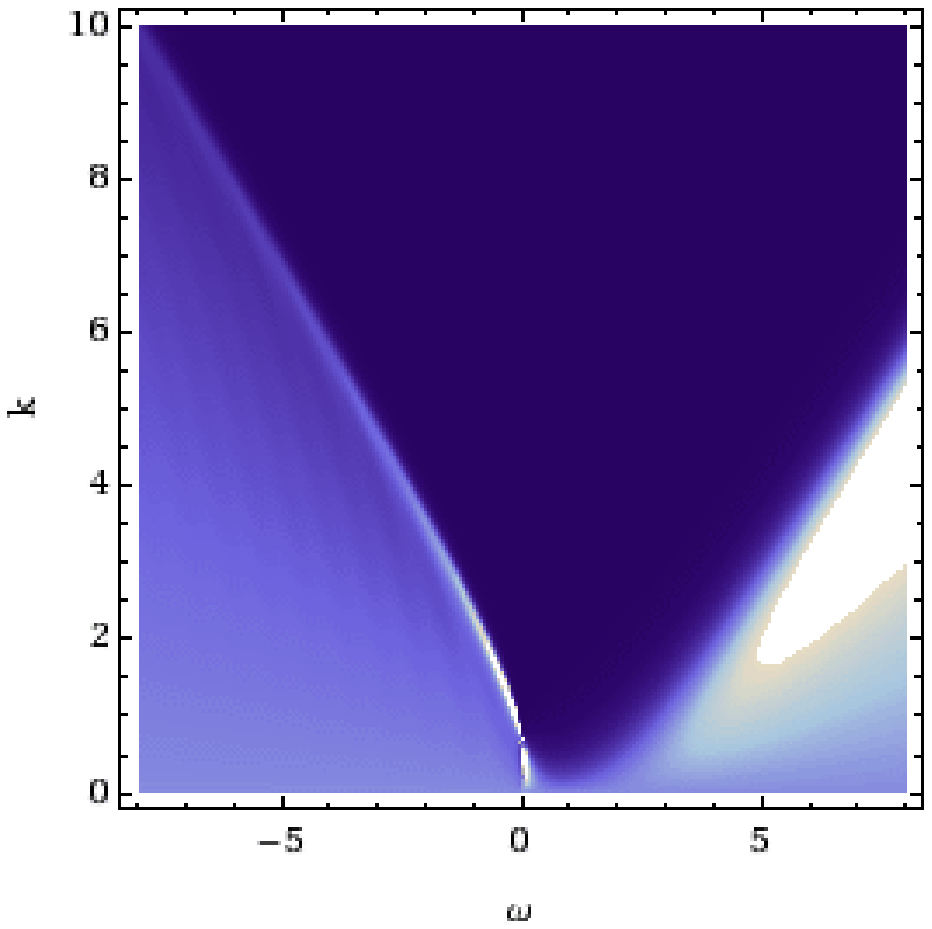}   
        \includegraphics[width=.48\textwidth]{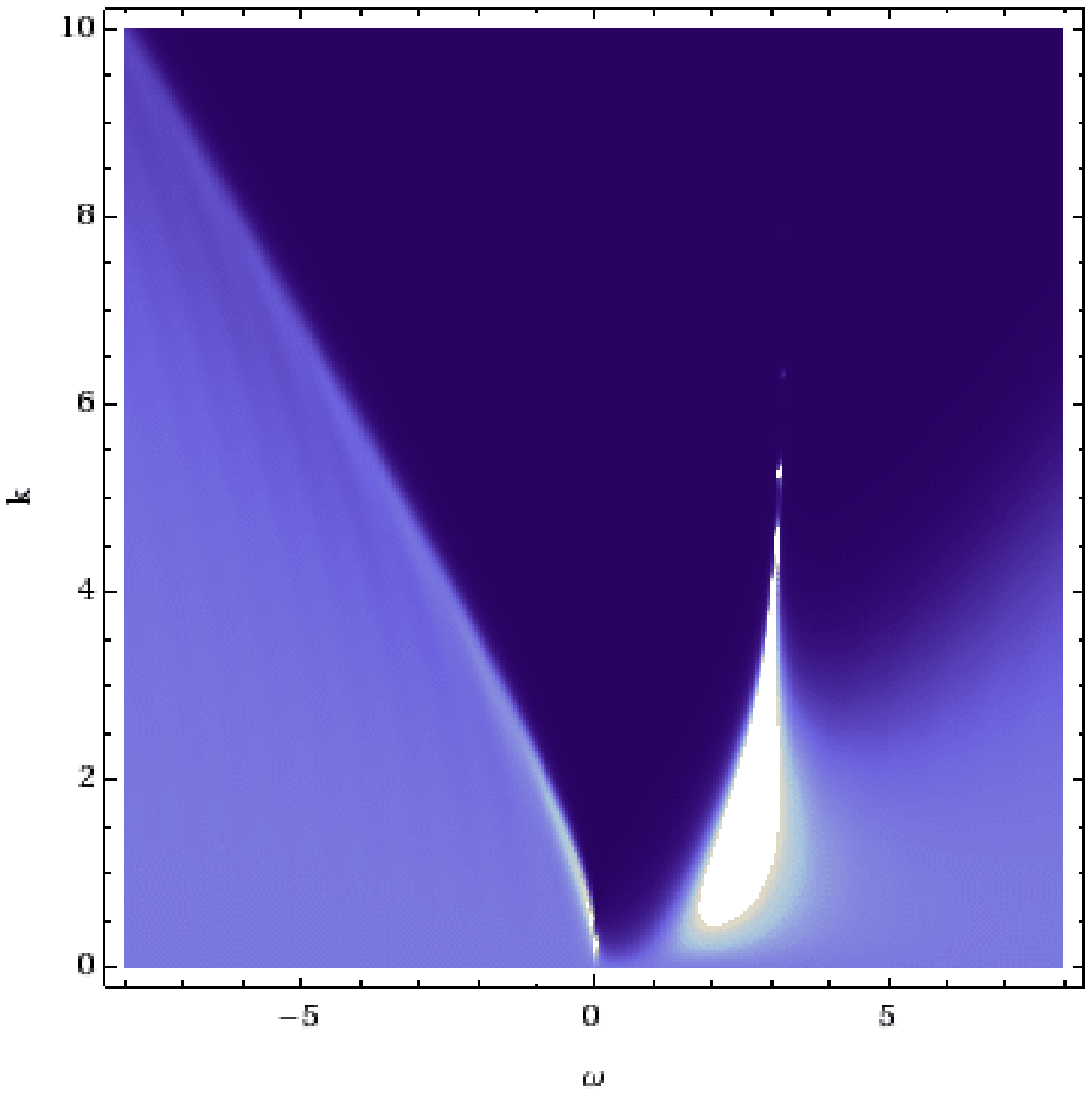}
      \includegraphics[width=.48\textwidth]{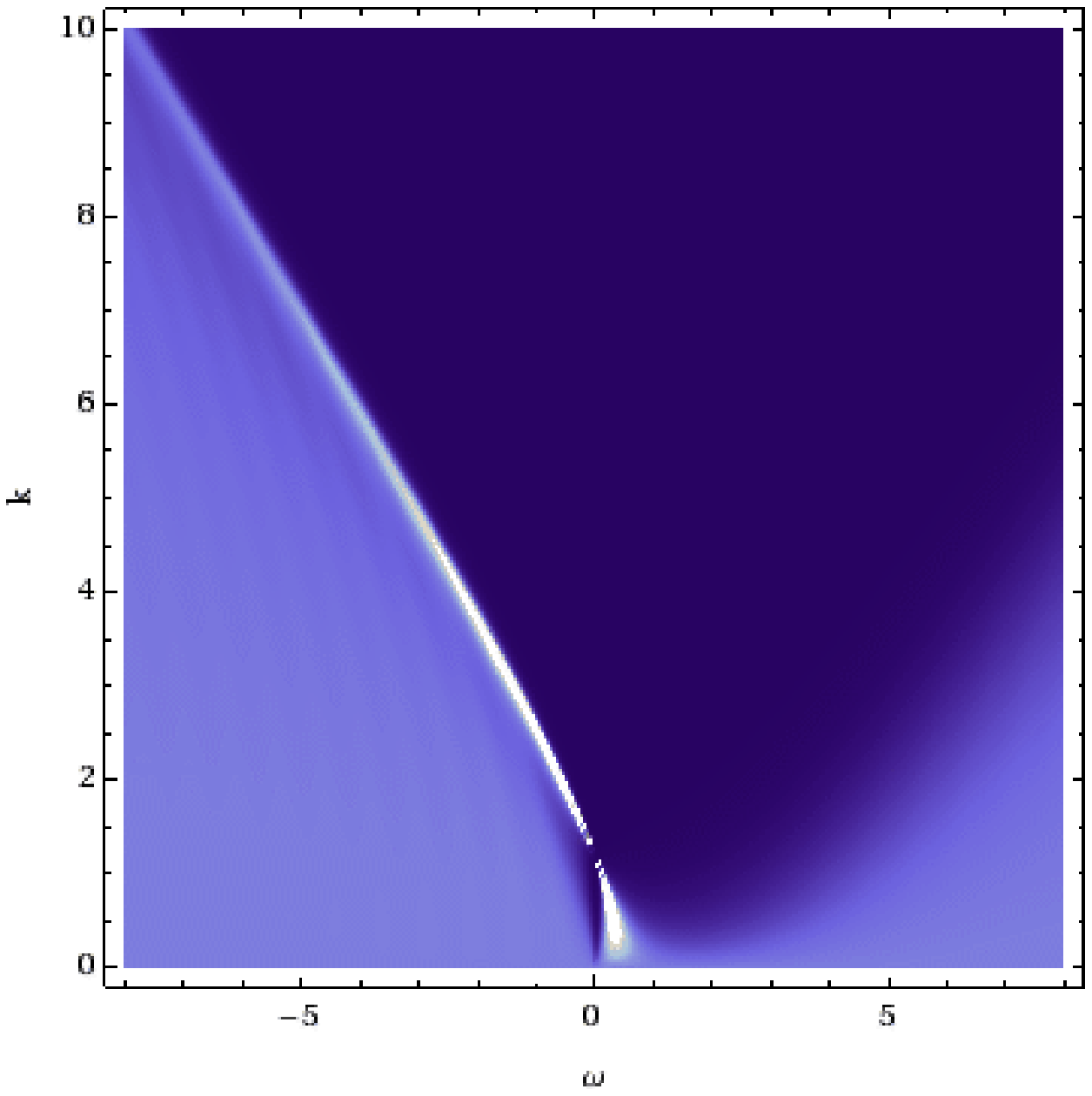}        \caption{The density plots of spectral function for the case of $q=2$, where the upstairs and downstairs plots are for the relativistic and non-relativistic fermionic fixed point respectively. }
       \label{q2}
\end{figure}

\begin{figure}[ht]
\centering      
\includegraphics[width=.48\textwidth]{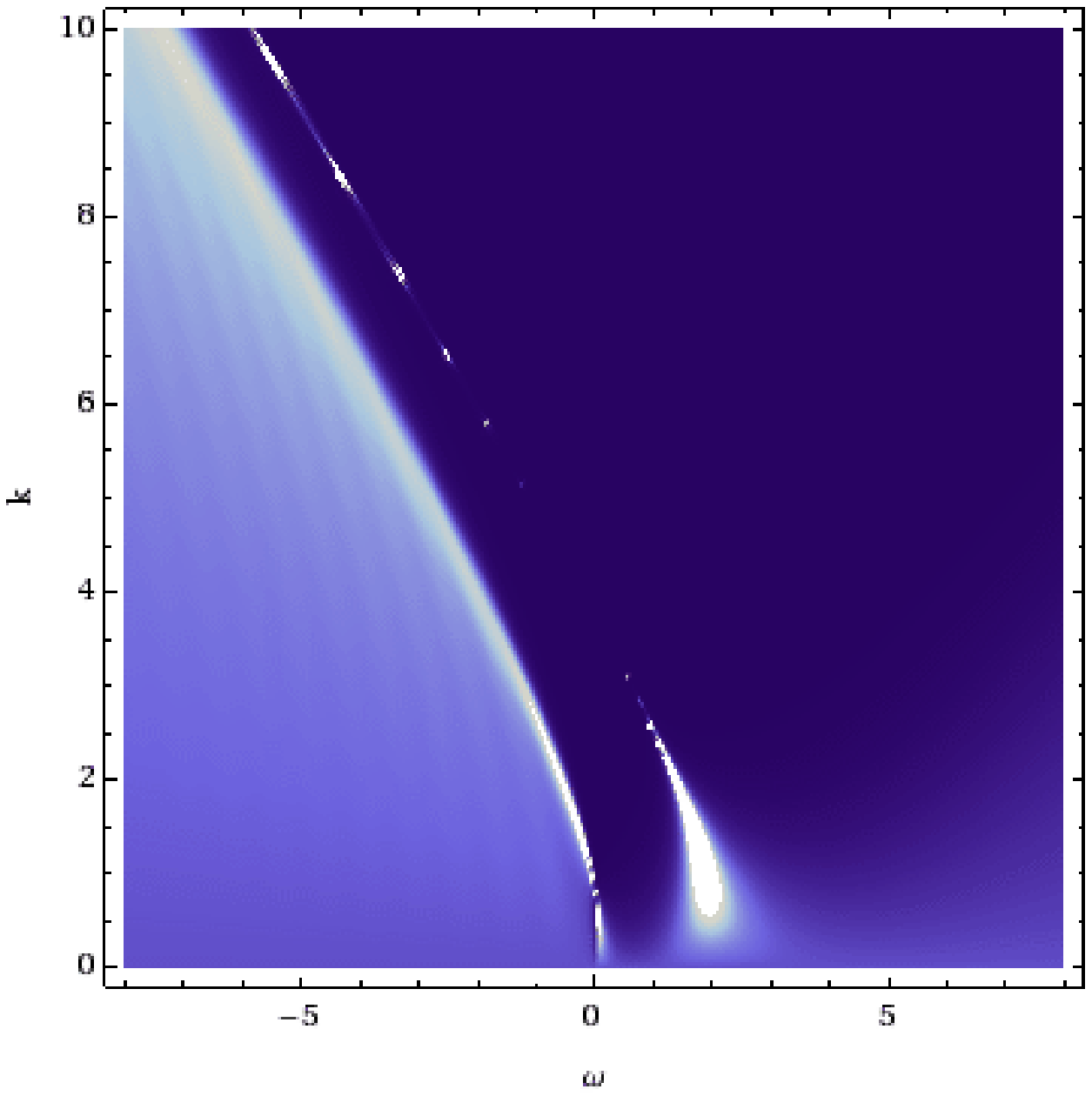}
      \includegraphics[width=.48\textwidth]{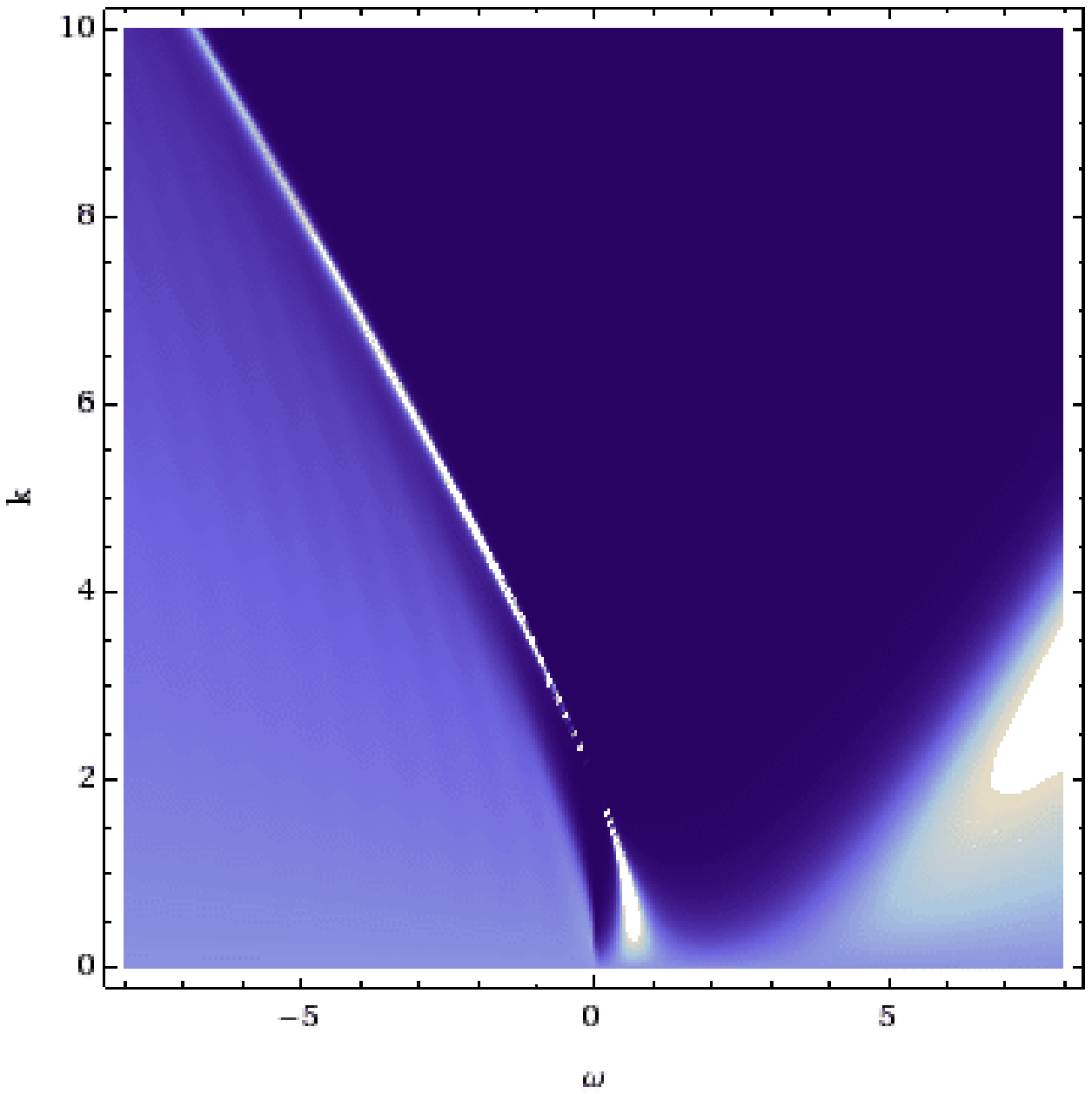}   
        \includegraphics[width=.48\textwidth]{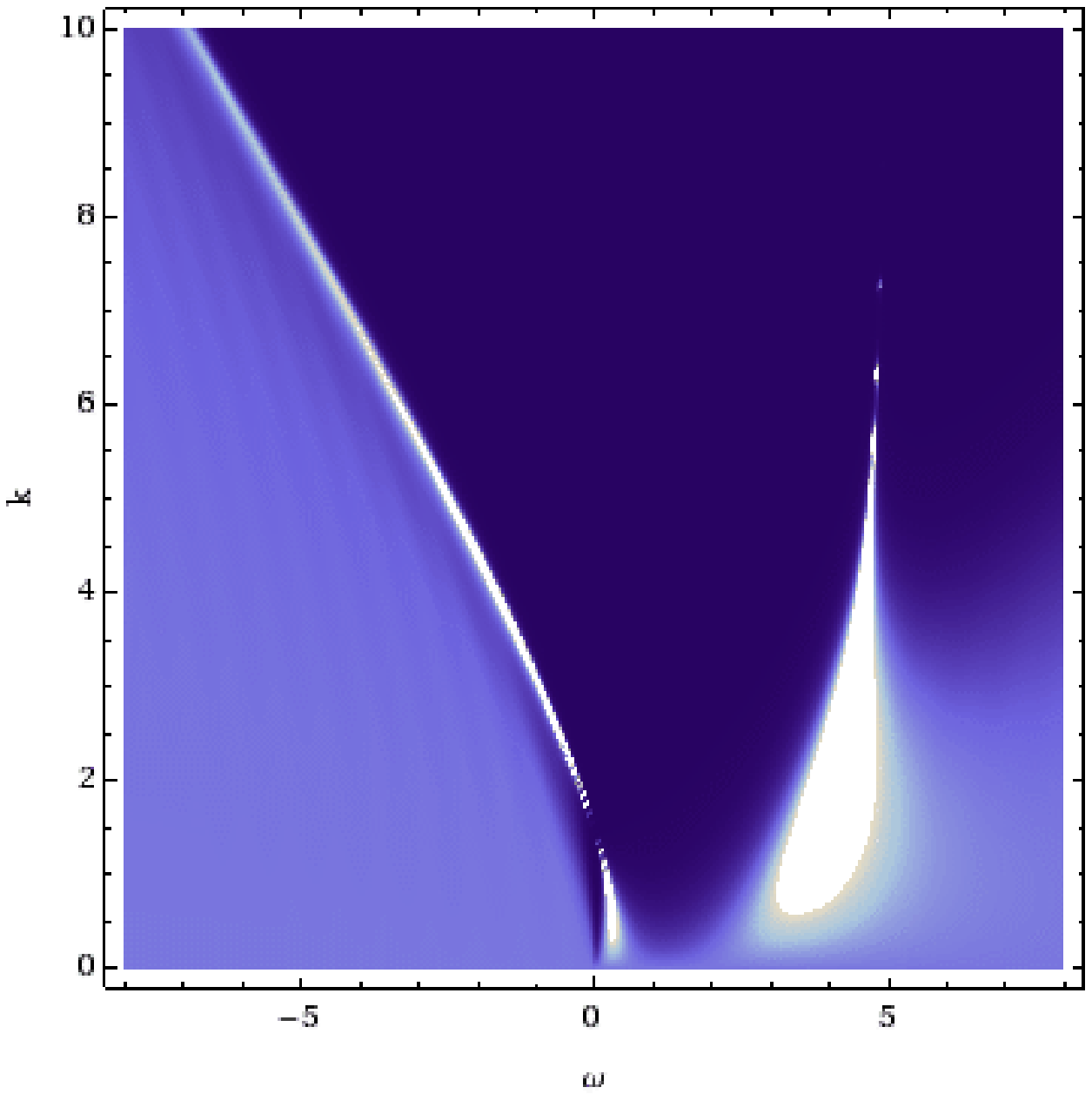}
      \includegraphics[width=.48\textwidth]{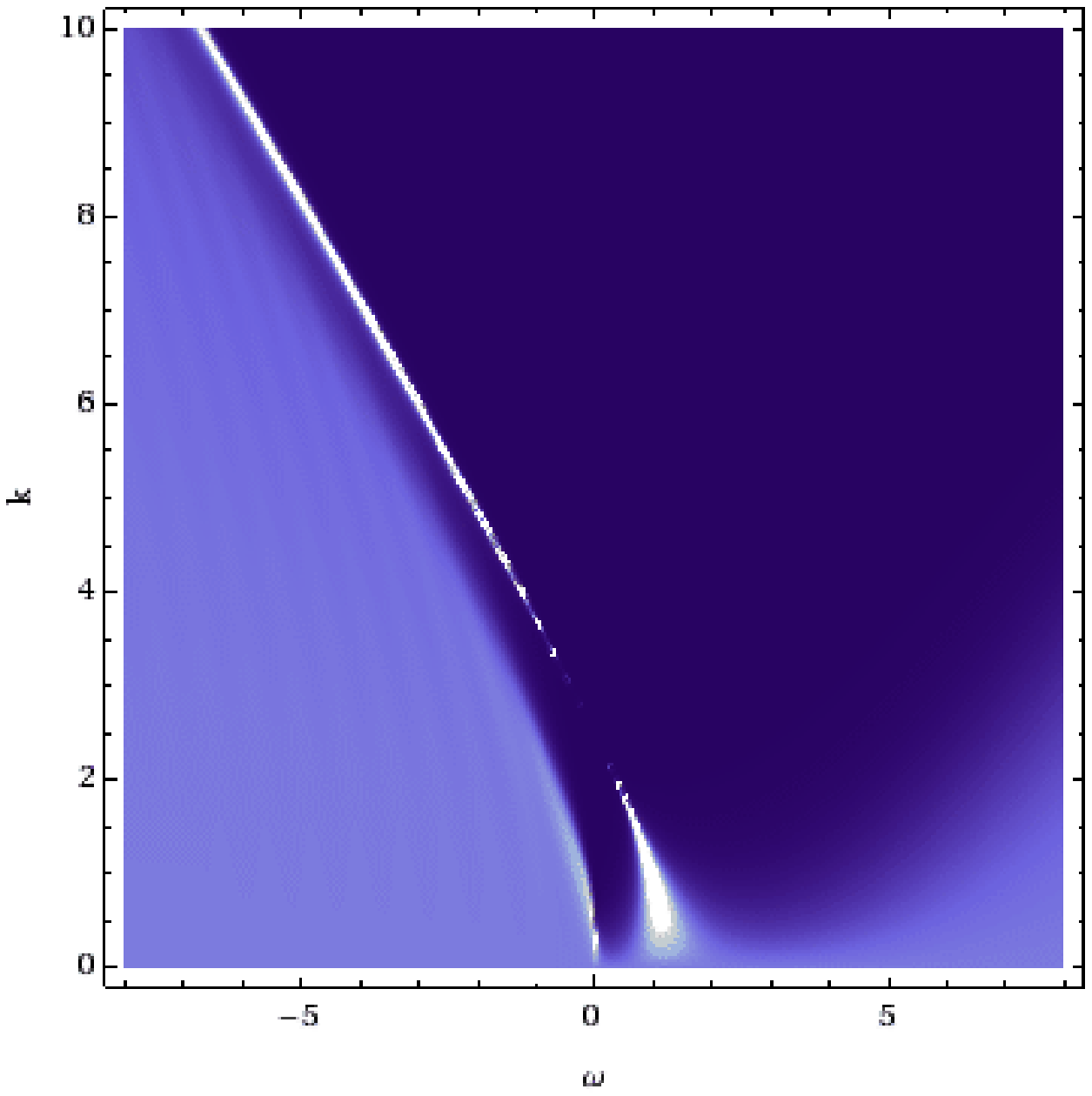}        \caption{The density plots of spectral function for the case of $q=3$, where the upstairs and downstairs plots are for the relativistic and non-relativistic fermionic fixed point respectively. }
       \label{q3}
\end{figure}

We would like to start presenting our numerical results by depicting the density plots of spectral function for $q=1, 2, 3$ in Figures \ref{q1}, \ref{q2}, \ref{q3} respectively, where the top left is for $\mathtt{Im}(G_1)$, the top right is for $\mathtt{Im}(G_2)$, the bottom left is for $\mathtt{Im}(\lambda_+)$, and the bottom right is for $\mathtt{Im}(\lambda_-)$.

\begin{figure}[ht]
\centering      
\includegraphics[width=.72\textwidth]{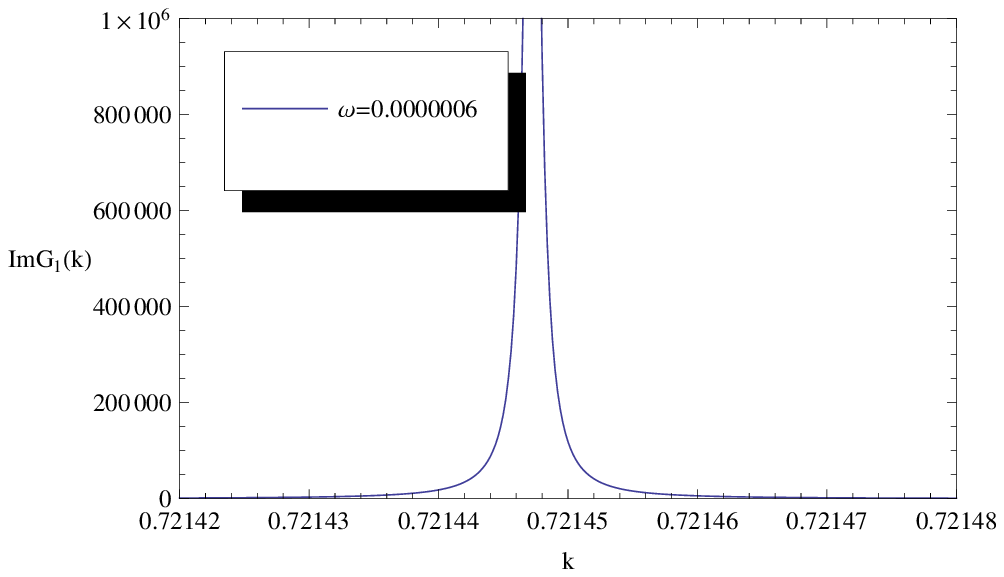}
            \caption{The Fermi momentum can be fixed to 0.721448 by setting $\omega=0.0000006$ for the case of $q=1$. }
       \label{fs}
\end{figure}

\begin{figure}[ht]
\centering      
\includegraphics[width=.72\textwidth]{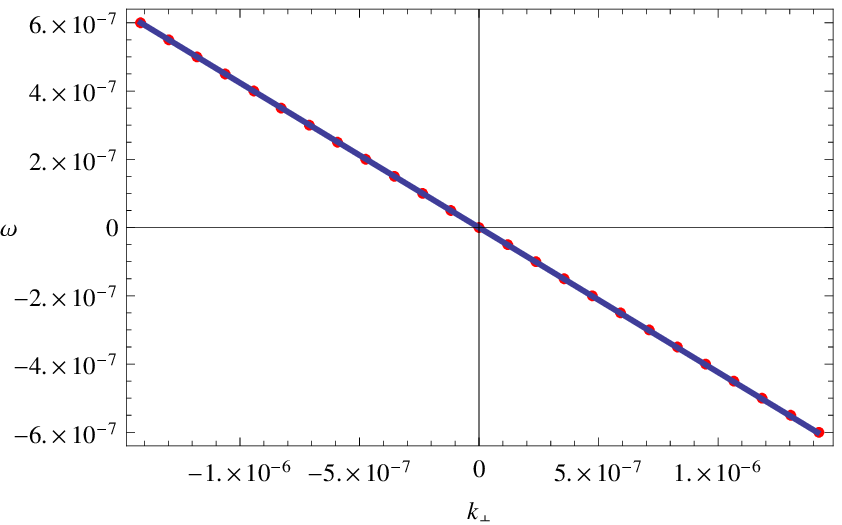}
            \caption{The dispersion relation around the Fermi surface at the relativistic fermionic fixed point in the case of $q=1$, where $k_\perp=k-k_F$, and the corresponding Fermi velocity can be further determined as $v_F=-0.424$. }
       \label{dr}
\end{figure}

 First,  in the presence of  the finite chemical potential the infinite flat band is always visible at the non-relativistic fixed point, but no longer exactly flat, namely gets mildly dispersed. In particular,  note that at the non-relativistic fermionic fixed point the soup in which the probe fermion is immersed remains relativistic, which implies that those modes with the momentum order of the chemical potential are mostly excited. Thus it is reasonable to see that the infinite band is destroyed or depleted from zero momentum up to the momentum order of the finite chemical potential by colliding the probe fermion with such modes in the relativistic soup. 
 On the other hand, as one goes to the higher and higher momentum,  the band recovers its flatness gradually with the width sharper and sharper, which arises because the high momentum modes sit outside of the light cone and the relativistic soup can not make them decay.  Furthermore, in the high momentum limit, the flat band is always shifted to $\omega=-\Omega$, which is conceivable as the frequency is measured relative to the chemical potential.

In addition, with the flat band standing at the non-relativistic fermionic fixed point, the formation of Fermi surface is apparently suppressed. Namely, one needs a larger value of charge parameter to make the Fermi surface show up at the non-relativistic fermionic fixed point,  compared to the situation happening at the relativistic fermionic fixed point. However, the presence of flat band appears not to change the low energy behavior around the Fermi surface in a drastic way. This arises because the flat band is somehow put by hand as sort of UV data while the low energy behavior is supposed to be essentially controlled by the near horizon geometry of our extremal charged dilatonic black hole, whatever it is.
\begin{table}[htbp] \centering
   \begin{tabular}{|c|c|c|}
   \hline
     &q=1(N)&q=1(R)\\
     \hline
     $k_F(v_F)$&no Fermi surface&0.721448(-0.424)\\
     \hline
         \end{tabular}

   \centering
   \begin{tabular}{|c|c|c|}
   \hline
     &q=2(N)&q=2(R)\\
     \hline
     $k_F(v_F)$&1.186305(-0.606)&0.708289(-0.292)\\
    &&2.195314(-0.794) \\
     \hline
       \end{tabular}

   \centering
   \begin{tabular}{|c|c|c|}
   \hline
     &q=3(N)&q=3(R)\\
     \hline
     $k_F(v_F)$&1.462513(-0.481)&0.810106(-0.267)\\
    &2.481857(-0.777)&1.976027(-0.639) \\
    &&3.749562(-0.867)     \\
    \hline
       \end{tabular}
    \caption{The Fermi momentum and Fermi velocity for both of the non-relativistic and relativistic fermionic fixed points in the case of $q=1, 2, 3$.}
   \label{fm}
\end{table}

\begin{figure}[ht]
\centering      
\includegraphics[width=.72\textwidth]{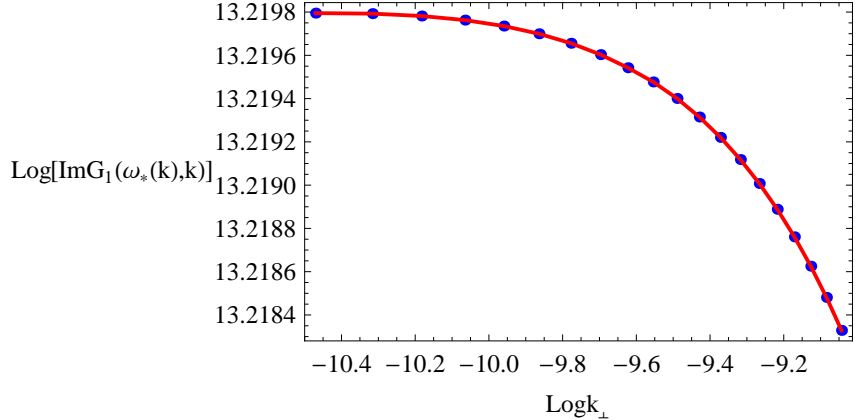}
            \caption{The scaling relation around the Fermi surface at the relativistic fermionic fixed point in the case of $q=1$, where $\omega_*(k)$ is the location of peak of spectral function at the given momentum $k$. }
       \label{scaling}
\end{figure}

It is noteworthy that the whole pattern documented above is qualitatively similar to that given by the probe fermion in the extremal Reissner-Nordstrom AdS black hole\cite{LT1}. But as alluded to above, due to the different near horizon geometry, the low energy behavior around the Fermi surface is bound to distinguish between them. To see this, now let us firstly identify the location of Fermi surface to the 6th digit, which, as illustrated in Figure \ref{fs},  can be achieved by using the fact that the location of peak of spectral function approaches the Fermi momentum as one takes the $\omega\rightarrow 0$ limit\footnote{Another way to pin down the position of Fermi surface is to use the fact the retarded Green function $G_R(\omega=0,k)$ develops a pole at the Fermi momentum, although the spectral function is automatically zero for $\omega=0$ because there is no source for the imaginary part of fermionic correlator due to the real boundary condition (\ref{zero}) and real flow equation (\ref{flow}).}. Then as shown in Figure \ref{dr}, we can further determine the dispersion relation around the Fermi surface, which turns out to be linear for all the cases we are considering here, not only irrespective of what kind of fermionic fixed point we work with, but also independent of the specific value of charge parameter we choose. We list the full result for the Fermi momentum and Fermi velocity in Table \ref{fm}. Such a linear dispersion means that the quasi-particle excitation is well defined and the probe sector has the linear specific heat as well, indicating that the dual liquid is like a Landau Fermi liquid, which is totally different from the situation given by the probe fermion in the extreme Reissner-Nordstrom AdS black hole\footnote{It is noteworthy that the probe sector does not always share the same thermodynamic behavior as the soup necessarily. Here the linear specific heat for the probe sector comes from the well known fact that the system consisting of the fermonic quasi-particle excitations has always the linear specific heat at low temperature\cite{Patheria}.}. In the latter case, for example, at the relativistic fermionic fixed point the dispersion relation goes like $\omega\propto k_\perp^p$, where $p$ decreases rapidly with increasing the charge parameter,  and only in the large charge limit, can the linear dispersion relation be achieved\cite{LMV}. However, to make sure whether the dual liquid is exactly of Landau Fermi type, we are left with one thing to check, namely to see whether the height of spectral function at the maximum scales as $k_\perp^{-w}$ with $w=1$ as $k_\perp\rightarrow 0$\cite{Senthil}. As demonstrated in Figure \ref{scaling}, $w$ turns out to be zero for the case of $q=1$, which can be checked to be also true for the other $q$ cases\footnote{Note that such a result is obviously at odds with the value of $w$ obtained in \cite{Wu}, where the only case of $q=2$ was considered at the relativistic fermionic fixed point and the corresponding scaling was given by $w=5$. Such a difference may come from the fact that the data used in \cite{Wu} were a little bit far away from the Fermi momentum. We are believed that our numerics is right. For one thing, our numerics reproduces the relevant results given by the probe fermion in the extremal Reissner-Nordstrom AdS black hole\cite{LMV}. For another, our numerical result is also consistent with the general argument $w\leq1$, which follows essentially from the requirement that the fermion occupation number should be always bounded around the Fermi surface\cite{Senthil}.}. So in this regard, the dual liquid does not behave as a Landau Fermi one.

\section{Conclusions}
We have investigated how the dual spectral function behaves at the non-relativistic as well as relativistic fermionic fixed point by working with the probe Dirac fermion in the extremal charged dilatonic black hole. As a result, although the whole pattern for the appearance of flat band as well as formation of Fermi surface is qualitatively similar to that extracted from the probe fermion in the extremal Reissner-Nordstrom AdS black hole, the low energy behavior around the Fermi surface exhibits some kind of universal behavior, which is totally different from what happens to the probe fermion in the extremal Reissner-Nordstrom AdS black hole. On one hand, the linear dispersion relation suggests the dual liquid is  a Fermi one. On the other hand, the scaling property of the maximal height of spectral function around the Fermi surface seems to indicate that the dual liquid does not behave exactly as a Landau Fermi liquid.

 We conclude with various generalizations worthy of further investigation.
 First, besides going beyond the massless case and heating up the dual soup to the finite temperature, it is interesting to investigate the probe fermion in other desirable dilatonic black holes such as those obtained in \cite{GKPT,CGKKM}, where the near horizon geometry has a favorable Lifshitz symmetry. In addition, as done in \cite{ELP,ELLP,GSW1,GSW2,LZ}, one is also tempted to investigate how the fermionic correlator is modified in the presence of the bulk dipole coupling. Last but not least, 
the universal low energy behavior we have found in this paper, especially irrespective of the specific value of charge parameter, begs a better understanding. Namely, one is required to see if this is also well captured by the so-called semi-holographic description\cite{FP}, or to be more precise, to see if the rigmarole of matching calculation can also been performed here such that the low energy behavior can be understood in an analytic way. We expect to explore these issues elsewhere.
 
 \section*{Acknowledgements}
Our thanks go to Nabil Iqbal, Fabio Rocha, and Kostas Skenderis  for their helpful discussions on black hole thermodynamics as well as David Vegh for his sharing the relevant Mathematica code.
WJL is grateful to Sijie Gao for his everlasting encouragement and relevant discussion. He is also indebted to  Jianpin Wu for sharing his experience in numerical calculations. RM wishes to thank the Max Planck Institute for Physics in Munich, as well as Theoretical Physics Institute at University of Jena for their hospitality during the final stage of this work. HZ would like to thank Zhi Wang for his useful discussion on the Landau Fermi liquid and non-Fermi liquid. He also likes to thank the organizers of Paris Meeting on Holography at Finite Density for their financial support and hospitality, where the relevant conversations with Joao Laia and David Tong are appreciated. WJL was supported in part by
the NSFC under grant No.10605006 together with 10975016 and by the
Fundamental Research Funds for the Central Universities.
RM and HZ were partially supported by a European Union grant FP7-REGPOT-2008-1-CreteHEPCosmo-228644. This research was also supported in part by the Project of Knowledge Innovation Program of Chinese Academy of Sciences, Grant No.KJCX2.YW.W10.

\end{document}